\documentclass[preprint,pteplogo]{ptephy_v1}

\usepackage{float}
\newcommand{\myvec}[1]{\mbox{\boldmath $#1$}}

\preprintnumber{2206-017} 

\title{Jammed Keplerian gas leads to the formation and disappearance of spiral arms in a coupled map lattice for astronomical objects}

\author{Erika Nozawa}
\affil{Department of Physics, Faculty of Core Research, Ochanomizu University, Tokyo 112-8610, Japan\email{papers@e-rika.net}}

\begin{document}

\begin{abstract}
The formation and disappearance of spiral arms are studied by focusing on jammed Keplerian gas
in a coupled map lattice (CML) with a minimal set of procedures for simulating diverse patterns in astronomical objects.
The CML shows that a spiral arm is a type of traffic jam,
and its motion is governed by both a gas inflow into and outflow from the jam.
In particular, we present a new approach to simply and directly evaluating the disappearance of spiral arms, called ``light-in and heavy-out''.
It is based on the gas flow rate difference between the light inflow and heavy outflow
leading to the disappearance of traffic jams.
Furthermore, we propose an approximate formula for the remaining lifetime of spiral arms,
which is immediately derived from the ``light-in and heavy-out'' approach without calculating their pattern speeds as in conventional differential rotation.
The proposed formula is successfully applied to the CML simulations.
\end{abstract}
\subjectindex{A31, A33} 
\maketitle

\section{Introduction}

Coupled map lattice (CML) is a useful dynamical system with discrete space and time, and continuous state variables \cite{Kaneko,Kanekos}.
It well reproduces experimental observations in spatially extended dynamical phenomena \cite{Yanagitab,Yanagitac,Yanagitad,Nishimori},
due to its simple construction and fast computation.
Furthermore, it offers new and suggestive perspectives on such phenomena
based on nonlinear dynamics.
Indeed, in the previous paper \cite{Nozawa} we have proposed a CML for astronomical objects to study the spiral pattern formation of accreting cold dense gas, such as seen in spiral galaxies \cite{GalaDy} or protoplanetary disks \cite{SEEDS},
and simulated dynamic patterns and properties agreeing with the several results of the conventional theories \cite{GalaDy,Lin-Shu} and observations \cite{Kuno,Miyoshi,Nadia,Kuno2}.
For example, the simulated spiral arms are pattern arms similar to density waves \cite{Lin-Shu}.
Additionally, we have also suggested in \cite{Nozawa} that the proposed CML can offer a new perspective on the dynamic properties of spiral arms, especially on the disappearance of spiral arms, based on the dynamics of traffic jams.

There are observations supporting that spiral arms are not material arms, but pattern arms such as density waves \cite{Kuno2,Laura}.
It suggests that there may be a simple and direct evaluation for the disappearance of spiral arms which takes into account the dynamic nature of patterns.
So far the disappearance of spiral arms has been discussed
based on the differential rotation in astrophysics \cite{GalaDy,Toomre,Meidt}.
However, the differential rotation was originally for material arms, not for pattern arms \cite{GalaDy}.
It seems to be a difficult and indirect evaluation for the disappearance of spiral arms to apply the differential rotation of material arms even to the pattern arms originally discussed in the framework of the rigid rotation, with calculation of their pattern speeds.
Let us recall that the motion of a pattern can be essentially different from that of the material in the pattern.
A well known example is the traffic jam of cars:
Cars (i.e., material) move forward although the traffic jam (i.e., pattern) moves backward,
and the traffic jam disappears naturally under a particular condition \cite{Nishinari}.

In this paper, we present a detailed analysis of the simulated spiral arms showing that they are astronomical traffic jams formed by jammed Keplerian gas around a central star,
and propose a new approach to simply and directly evaluating the disappearance of spiral arms in terms of the ``light-in and heavy-out'',
based on the CML suggestions.
The ``light-in and heavy-out'' is the gas flow rate difference between a light (i.e., low density) inflow into and heavy (i.e., high density) outflow from the jam.
This gas flow rate difference decreases the mass of the jam, and leads to its disappearance.
The ``light-in and heavy-out'' approach is consistent with the well known fact in astrophysics that
gas clumps flow into a spiral arm, and they are compressed and finally flow out as massive gas clumps like stars \cite{GalaDy} (or possibly even black holes \cite{Rana}).

We also propose an approximate formula for the remaining lifetime of spiral arms from the ``light-in and heavy-out'' approach.
It is immediately given by dividing the mass of traffic jams by the gas flow rate difference between the inflow and outflow, without calculating the pattern speeds of spiral arms.
The approximate values obtained are consistent with the remaining lifetimes in the CML simulations.
We expect that this approximate formula can be generally applied to the observational data of spiral arms
in spiral galaxies or protoplanetary disks
by assuming that spiral arms are of the nature of traffic jams.
This is because it would take a great deal of effort to make the observations to obtain the pattern speeds of spiral arms.

The present paper is organized as follows.
In Section \ref{model}, we explain briefly a CML for astronomical objects, which we have introduced in \cite{Nozawa}.
In Section \ref{Dynamic properties of a spiral arm}, we study the dynamic properties of a simulated spiral arm from three aspects in traffic jams.
First, the formation of the spiral arm occurs due to the jammed motion of Keplerian gas passing through the arm.
Second, the motion of the spiral arm is given as the movement of the traffic jam.
Third, the disappearance of the spiral arm and especially its remaining lifetime are simply and directly evaluated from the gas flow rate difference between the light inflow into and heavy outflow from the jam, which is called the ``light-in and heavy-out''.
Summary and discussion are given in Section \ref{summary}.

\section{Model}
\label{model}

We briefly review a coupled map lattice (CML) for simulating diverse patterns, especially spiral patterns, in astronomical objects \cite{Nozawa}.
The CML consists of a minimal set of procedures: gravitational interaction and advection.

We study a system of accreting cold dense gas moving on a two-dimensional region,
such as an accretion disk well known in active galactic nuclei and protoplanetary disks \cite{GalaDy}.
The region is divided into $N_{x}\times N_{y}$ square cells.
The gas in each cell forms a macroscopic gas clump through the collision of dense gas \cite{GalaDy},
which is treated as an assembly of virtual gas particles (not gas molecules).
For simplicity, we assume that gas particles in each cell are distributed uniformly and carried by the same flow.

We introduce a finite two-dimensional square lattice by placing a lattice point on the center of each cell.
The distance between the nearest neighbor lattice points is set to one, and each cell size is also one.
The lattice points are labeled $ij$ ($i=0,1,\cdots,N_{x}-1$ and $j=0,1,\cdots,N_{y}-1$)
and their positions are given by the position vectors $\myvec{r}_{ij}=(i,j)=i\myvec{e}_{x}+j\myvec{e}_{y}$,
where $\myvec{e}_{x}$ and $\myvec{e}_{y}$ are unit vectors parallel to the $x$- and $y$-axis respectively.

We define two kinds of field variables,
the mass $m_{ij}^{t}$ and the velocity $\myvec{v}_{ij}^{t}=v_{x\, ij}^{t}\myvec{e}_{x}+v_{y\, ij}^{t}\myvec{e}_{y}$
of the gas clump at lattice point $ij$ at discrete time $t$.
They are coarse-grained variables given by the total mass and the flow of gas particles in the cell at lattice point $ij$, respectively.

We decompose the spiral pattern formation in astronomical objects into two important elementary processes, a gravitational interaction process and an advection process.
In the former process,
the gas clump velocity $\myvec{v}_{ij}^{t}$ is changed by the gravitational interaction among gas clumps.
In the latter process,
the gas clump mass $m_{ij}^{t}$ and velocity $\myvec{v}_{ij}^{t}$ are changed
since gas particles move and collide to form new gas clumps
by the flows, i.e., the gas clump velocities resulting from the former process.

We formulate the gravitational interaction process as an Eulerian procedure \cite{Kaneko,Yanagitac} $T_{g}$ in the lattice picture \cite{Nozawa}.
In the procedure $T_{g}$,
the gas clump velocity $\myvec{v}_{ij}^{t}$ is changed to $\myvec{v}_{ij}^{*}$ due to the gravitational interaction from the other gas clumps at lattice points $kl$,
where $*$ represents an intermediate time between discrete times $t$ and $t+1$.
The procedure $T_{g}$ is defined by the following maps:
\begin{equation}
\label{eqn:Tg_m}
m_{ij}^{*}=m_{ij}^{t},
\end{equation}
\begin{equation}
\label{eqn:Tg_v}
\myvec{v}_{ij}^{*}=\myvec{v}_{ij}^{t}
-\gamma\tau_{g}\sum_{k=0}^{N_{\mathstrut x}-1}\sum_{l=0}^{N_{\mathstrut y}-1}
\frac{(1-\delta_{ik}\delta_{jl})m_{kl}^{t}}{|\myvec{r}_{ij}-\myvec{r}_{kl}|^{2}}
\frac{\myvec{r}_{ij}-\myvec{r}_{kl}}{|\myvec{r}_{ij}-\myvec{r}_{kl}|},
\end{equation}
where $\gamma$ is the gravitational constant,
$\tau_{g}$ the time interval for the procedure $T_{g}$
and $\delta$ the Kronecker delta.
As shown in Eq.\,\ref{eqn:Tg_m},
the gas clump mass $m_{ij}^{t}$ does not change in the procedure $T_{g}$.
The procedure $T_{g}$ of Eqs.\,\ref{eqn:Tg_m} and \ref{eqn:Tg_v} has a computational cost of $O(N^{2})$, where $N$ is the total number of lattice points.

We formulate the advection process as a Lagrangian procedure \cite{Kaneko,Yanagitac} $T_{a}$ in the particle picture \cite{Nozawa}.
In the procedure $T_{a}$,
each flow $\myvec{v}_{kl}^{*}$ resulting from the procedure $T_{g}$
carries gas particles with their total mass $m_{kl}^{*}$ and momentum $m_{kl}^{*}\myvec{v}_{kl}^{*}$
from the cell at lattice point $kl$
to a cell-sized area centered at the position
\begin{equation}
\label{eqn:kl}
(\tilde{k},\tilde{l})=(k+v_{x\, kl}^{*}\tau_{a},l+v_{y\, kl}^{*}\tau_{a}),
\end{equation}
where $\tau_{a}$ is the time interval for the procedure $T_{a}$.
When the cell-sized areas overlap the cell at lattice point $ij$,
the size $w_{ijkl}^{*}$ of each overlap area is given by \cite{Nozawa}
\begin{equation}
\label{eqn:w_ijkl}
w_{ijkl}^{*}=
\left(\delta_{i\lfloor \tilde{k}\rfloor}\delta_{j\lfloor \tilde{l}\rfloor}
+\delta_{i\lfloor \tilde{k}\rfloor +1}\delta_{j\lfloor \tilde{l}\rfloor}
+\delta_{i\lfloor \tilde{k}\rfloor +1}\delta_{j\lfloor \tilde{l}\rfloor +1}
+\delta_{i\lfloor \tilde{k}\rfloor}\delta_{j\lfloor \tilde{l}\rfloor +1}\right)
\left(1-\left|\tilde{k}-i\right|\right)\left(1-\left|\tilde{l}-j\right|\right),
\end{equation}
where $\lfloor \bullet \rfloor$ is the floor function.
In the cell at lattice point $ij$,
gas particles in the overlap areas collide with each other and are mixed into one clump.
This collision and mixture lead to
the formation of a new gas clump whose mass and velocity are $m_{ij}^{t+1}$ and $\myvec{v}_{ij}^{t+1}$ respectively.
Thus, the advection procedure $T_{a}$ is defined by the following maps:
\begin{equation}
\label{eqn:Ta_m}
m_{ij}^{t+1}=\sum_{k=0}^{N_{\mathstrut x}-1}\sum_{l=0}^{N_{\mathstrut y}-1}w_{ijkl}^{*}m_{kl}^{*},
\end{equation}
\begin{equation}
\label{eqn:Ta_v}
\myvec{v}_{ij}^{t+1}=\frac{1}{m_{ij}^{t+1}}\sum_{k=0}^{N_{\mathstrut x}-1}\sum_{l=0}^{N_{\mathstrut y}-1}
w_{ijkl}^{*}m_{kl}^{*}\myvec{v}_{kl}^{*}.
\end{equation}
In Eq.\,\ref{eqn:Ta_v}, when gas clump mass $m_{ij}^{t+1}$ vanishes,
gas clump velocity $\myvec{v}_{ij}^{t+1}$ is set to be also zero.
The procedure $T_{a}$ of Eqs.\,\ref{eqn:Ta_m} and \ref{eqn:Ta_v} has a low computational cost of $O(N)$, as shown in \cite{Nozawa}.

We construct the time evolution of gas clump mass $m_{ij}^{t}$ and velocity $\myvec{v}_{ij}^{t}$ for one step (from discrete time $t$ to $t+1$) by the following successive operations of the gravitational interaction procedure $T_{g}$ and the advection procedure $T_{a}$:
\begin{equation}
\label{eqn:dynamics}
\left(
\begin{array}{c}
{m_{ij}^{t}\rule{0pt}{10pt}} \\
{\myvec{v}_{ij}^{t}\rule{0pt}{10pt}} \\
\end{array}
\right)
\stackrel{T_{g}}{\longmapsto}
\left(
\begin{array}{c}
{m_{ij}^{*}\rule{0pt}{10pt}} \\
{\myvec{v}_{ij}^{*}\rule{0pt}{10pt}} \\
\end{array}
\right)
\stackrel{T_{a}}{\longmapsto}
\left(
\begin{array}{c}
{m_{ij}^{t+1}\rule{0pt}{10pt}} \\
{\myvec{v}_{ij}^{t+1}\rule{0pt}{10pt}} \\
\end{array}
\right).
\end{equation}
On the time evolution of Eq.\,\ref{eqn:dynamics},
the total mass $\sum_{i,j}m_{ij}^{t}$, total momentum $\sum_{i,j}m_{ij}^{t}\myvec{v}_{ij}^{t}$ and total angular momentum $\sum_{i,j}\myvec{r}_{ij}\times m_{ij}^{t}\myvec{v}_{ij}^{t}$ of the system are conserved \cite{Nozawa}.

The simulations were performed according to the following settings:
Lattice size $N_{x}\times N_{y}$ is $50\times 50$;
Gravitational constant $\gamma$ is one and time intervals $\tau_{g}$ and $\tau_{a}$ one;
The initial gas clump mass $m_{ij}^{0}$ is given by a uniform random number within $[0,2/(N_{x}N_{y})]$;
The initial gas clump velocity $\myvec{v}_{ij}^{0}$ is given by zero;
The boundary conditions are open.
In this settings, we observed that the initial nonstationary state is relaxed
through the formation of a central star,
and after that, grand design spiral patterns arise spontaneously
due to the chaotic gas ejection from the central star \cite{Nozawa}.

Under the same initial total mass, the formation and disappearance of spiral arms as presented below were also observed when the lattice size $N_{x}\times N_{y}$ is $30\times 30$, $\cdots$, $100\times 100$, and when using more stationary initial conditions (i.e., $\myvec{v}_{ij}^{0}\neq\myvec{0}$) such as a slowly moving and rotating central star with Keplerian gas.
The total mass, total momentum and total angular momentum are not conserved completely in the simulations,
since gas particles move out through the boundary of the finite lattice.
We note that the mass lost is a very small quantity ($0.7\%$ of the initial total mass)
through the following simulation from $t=0$ to $t=1520$.

\section{Dynamic properties of a spiral arm based on traffic jams}
\label{Dynamic properties of a spiral arm}

\begin{figure}[htbp]
  \begin{center}
    \includegraphics[scale=0.996]{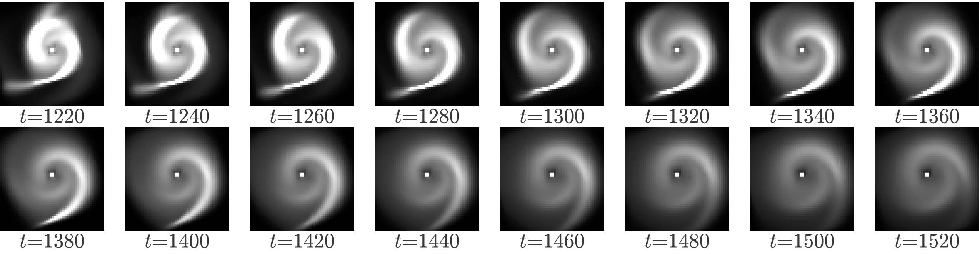}
  \end{center}
  \caption{
	Snapshots of gas clump masses $m_{ij}^{t}$.
  The brightness of cells represents the gas clump mass $m_{ij}^{t}$
  ($i=0,1,\cdots,49$, $j=0,1,\cdots,49$ and $t=1220,1240,\cdots,1520$)
  in the range of 0 to $1.2\times 10^{-6}$.
  }
  \label{fig:lmf.eps}
\end{figure}

The formation and disappearance of spiral patterns is shown in a series of snapshots in Fig.1. It takes only 18 sec to obtain this result with a personal computer by using the fast computation of the CML for astronomical objects.
The upper row snapshots at $t=1220,1240,\cdots,1360$ roughly show the formation of spiral arms,
and the lower row at $t=1380,1400,\cdots,1520$ their disappearance.
Each snapshot has $50\times 50$ cells, and the brightness of the cells represents the gas clump mass $m_{ij}^{t}$.

In the simulation, the contracting central star ejects gas particles chaotically from $t=990$ to $t=995$,
and they induce the traffic jams of Keplerian gas particles around the central star.
The jammed Keplerian gas particles form a pair of small newborn spiral arms up to $t=1000$
and thus a grand design spiral pattern (that is, two-arm spiral pattern) appears and becomes larger, as in the snapshots at $t=1220,1240,\cdots,$ and $1360$.
After that, the spiral pattern disappears gradually with the disappearance of traffic jams up to $t=1528$, as in the snapshots at $t=1380,1400,\cdots,$ and $1520$.
The supplementary movie of the simulation from $t=980$ to $t=1520$ is also available.
Hereafter we focus on the time around $t=1360$, the turning point from the formation to disappearance of spiral arms.

\subsection{Motion of gas particles crossing a spiral arm}
\label{Motion of gas particles crossing a spiral arm}

\begin{figure}[htbp]
  \begin{center}
    \includegraphics[scale=0.88]{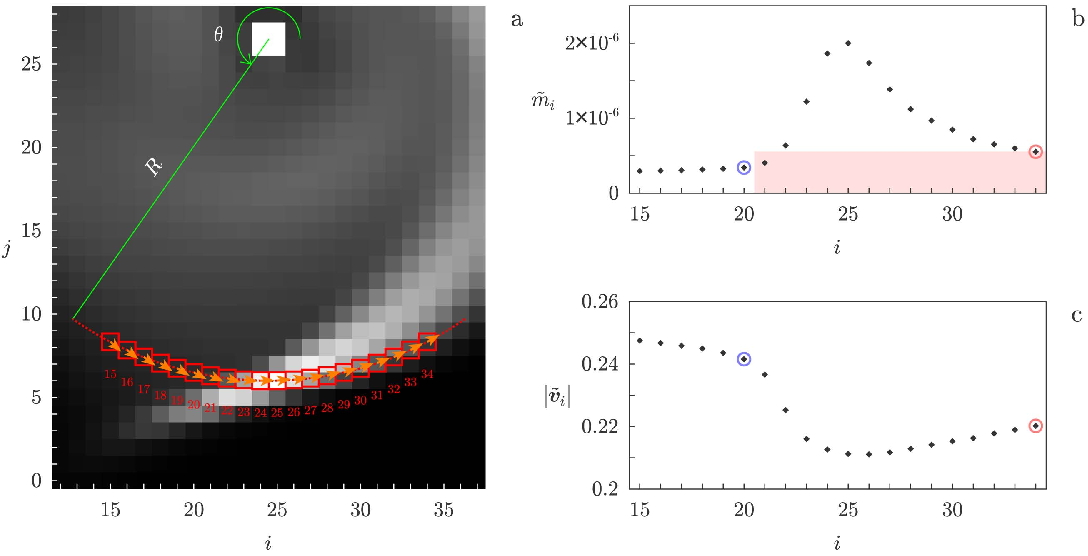}
  \end{center}
  \caption{
	Motion of gas particles crossing the arm along the circular flow around the central star.
	(a) Circular flow around the central star.
	The brightness of cells represents gas clump masses $m_{ij}^{t}$ in the range of 0 to $2\times 10^{-6}$,
	the red dotted line the circular flow,
	the red squares rearranged cells $i$ ($i=15,16,\cdots,34$)
	and the orange arrows average gas clump velocities $\tilde{\myvec{v}}_{i}$.
	(b) Average gas clump density $\tilde{m}_{i}$.
	(c) Average gas clump speed $|\tilde{\myvec{v}}_{i}|$.
  }
  \label{fig:cm.eps}
\end{figure}

Gas particles move along Keplerian flows around the central star,
and these Keplerian gas particles become jammed while crossing a spiral arm.
The red dotted line in Fig.\,\ref{fig:cm.eps}a shows a Keplerian flow which crosses the spiral arm from the left to the right at $t=1360$.
Figs.\,\ref{fig:cm.eps}b and \ref{fig:cm.eps}c show the density and speed changes of gas particles along the Keplerian flow, respectively.
The results in Figs.\,\ref{fig:cm.eps}b and \ref{fig:cm.eps}c are obtained through the following three steps.

First, we define the Keplerian flow approximated
as a circular flow at radial distance $R$ from the center of gravity $\myvec{r}_{S}$ of the central star.
As shown in Fig.\,\ref{fig:cm.eps}a,
the distance $R$ is determined with the Keplerian flow crossing the most dense part of the spiral arm,
which is located at the lattice points $24\ 6$ and $25\ 6$.
Here the central star is located at the lattice points $24\ 26$, $25\ 26$, $24\ 27$ and $25\ 27$,
and the center of gravity $\myvec{r}_{S}$ is thus given by
\begin{equation}
\label{eqn:rSt}
\myvec{r}_{S}
=(x_{S},y_{S})
=\frac{\sum\limits_{i=24}^{25}\sum\limits_{j=26}^{27}m_{ij}^{t}\myvec{r}_{ij}}
{\sum\limits_{i=24}^{25}\sum\limits_{j=26}^{27}m_{ij}^{t}},
\end{equation}
and the distance $R$ is given by
\begin{equation}
\label{eqn:R}
R=y_{S}-6.
\end{equation}

Second, we rearrange cells (the red squares in Fig.\,\ref{fig:cm.eps}a) along the Keplerian flow.
The position vectors $\myvec{r}_{i}$ of rearranged cells $i$ ($i=15,16,\cdots,34$) are given by
\begin{equation}
\label{eqn:ri}
\myvec{r}_{i}=\left(i,y_{S}+R\sin\theta (i)\right),
\end{equation}
where $\theta (i)$ is an angle in the polar coordinate with pole $\myvec{r}_{S}$ and defined by
\begin{equation}
\label{eqn:theta}
\theta (i)=\cos^{-1}\left(\frac{i-x_{S}}{R}\right)\quad (\pi\le\theta (i)\le 2\pi).
\end{equation}

Third, using area-weighted average (see Appendix \ref{Area-weighted average of field variables}),
we obtain the average gas density $\tilde{m}_{i}=\tilde{m}^{t}(\myvec{r}_{i})/1$ with the cell size 1,
the average gas velocity $\tilde{\myvec{v}}_{i}=\tilde{\myvec{v}}^{t}(\myvec{r}_{i})$ (the orange arrows in Fig.\,\ref{fig:cm.eps}a),
and the average gas speed $|\tilde{\myvec{v}}_{i}|$ in rearranged cell $i$.
Figs.\,\ref{fig:cm.eps}b and \ref{fig:cm.eps}c
show the average gas density $\tilde{m}_{i}$ and speed $|\tilde{\myvec{v}}_{i}|$ respectively.

Fig.\,\ref{fig:cm.eps} clearly shows the jammed motion of Keplerian gas particles in the arm.
While crossing the arm, Keplerian gas particles change their motion as follows:
(1) When they flow into the arm ($i=20,\cdots,24$ in Fig.\,\ref{fig:cm.eps}a),
density $\tilde{m}_{i}$ increases rapidly
(from $3.4\times 10^{-7}$ to $1.9\times 10^{-6}$ in Fig.\,\ref{fig:cm.eps}b)
and thus speed $|\tilde{\myvec{v}}_{i}|$ is decelerated quickly
(from 0.24 to 0.21 in Fig.\,\ref{fig:cm.eps}c);
(2) In the arm ($i=24,\cdots,27$ in Fig.\,\ref{fig:cm.eps}a),
they are jammed, that is, keep high density $\tilde{m}_{i}$
(in the range of $1.4\times 10^{-6}$ to $2.0\times 10^{-6}$ in Fig.\,\ref{fig:cm.eps}b)
and move with lower speed $|\tilde{\myvec{v}}_{i}|$
(0.21 in Fig.\,\ref{fig:cm.eps}c);
(3) When they flow out of the arm ($i=27,\cdots,34$ in Fig.\,\ref{fig:cm.eps}a),
density $\tilde{m}_{i}$ decreases slowly
(from $1.4\times 10^{-6}$ to $5.5\times 10^{-7}$ in Fig.\,\ref{fig:cm.eps}b)
and thus speed $|\tilde{\myvec{v}}_{i}|$ is accelerated gradually
(from 0.21 to 0.22 in Fig.\,\ref{fig:cm.eps}c).
Here we note that the jammed motion of gas particles also occurs in Keplerian flows at different radial distance $R$ and different time $t$.

\subsection{Motion of a spiral arm}
\label{Motion of a spiral arm}

\begin{figure}[htbp]
  \begin{center}
    \includegraphics[scale=0.53]{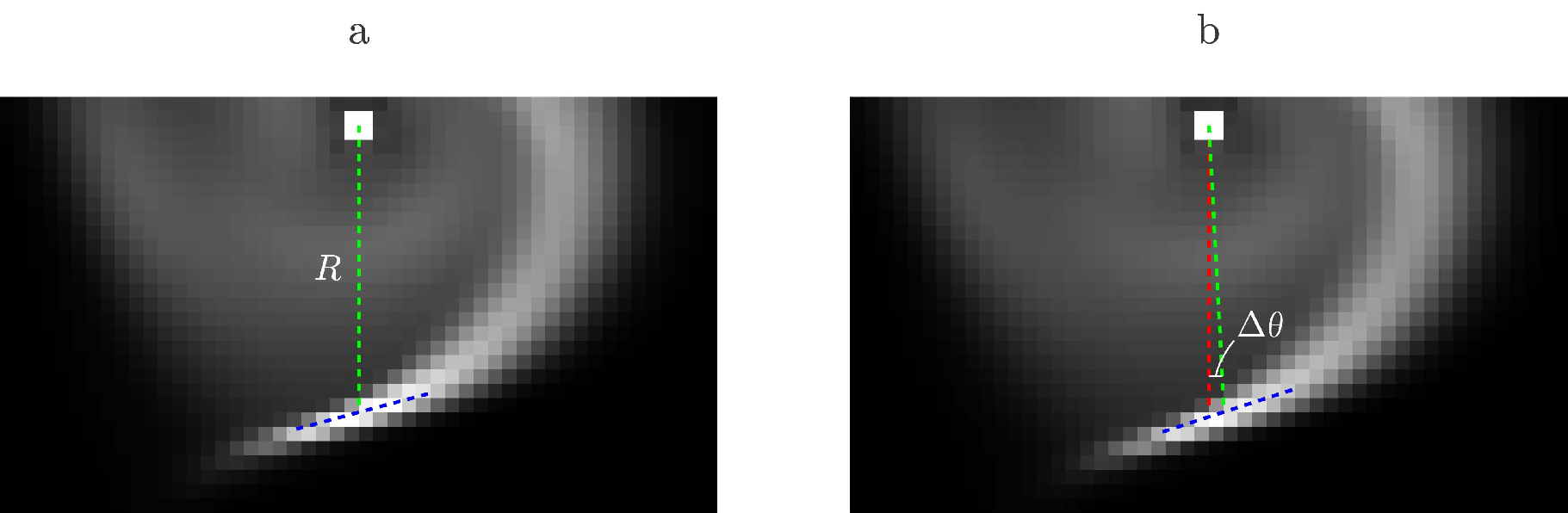}
  \end{center}
  \caption{
	Rotation of the spiral arm.
  (a) Snapshot of gas clump masses $m_{ij}^{t}$ at $t=1350$. (b) Same, but for $t=1360$.
  The green dotted lines are drawn from $\myvec{r}_{S}$ to the center of the blue dotted lines parallel to the arm.
  The green dotted line at $t=1360$ is rotated by $\Delta\theta$ from the red dotted line.
  }
  \label{fig:pv_1350_1360.eps}
\end{figure}

There is an astronomical traffic jam formed by high dense jammed Keplerian gas particles in a spiral arm, as described in the previous subsection.
The traffic jam moves along the circular flow not only with the velocity $v_{g}$ of jammed Keplerian gas particles,
but also with the velocity $v_{io}$ caused by both a gas inflow into and outflow from the jam.
This circular movement of the traffic jam is observed as the rotation of the spiral arm with the velocity
\begin{equation}
\label{eqn:vp}
v_{j}=v_{g}+v_{io}.
\end{equation}

The velocity $v_{io}$ is opposite to the velocity $v_{g}$, which is well known as a property of traffic jams \cite{Nishinari}.
We briefly explain the opposite velocity $v_{io}$.
In the tail of the traffic jam (cells $i\le 23$ in Fig.\,\ref{fig:cm.eps}),
Keplerian gas particles are decelerated to be part of the jam,
and thus the tail of the jam becomes longer.
On the other hand,
in the front of the jam (cells $i\ge 27$ in Fig.\,\ref{fig:cm.eps}),
jammed Keplerian gas particles are accelerated to leave the jam,
and thus the front of the jam becomes shorter.
The longer tail and shorter front leads to the backward movement of the jam with the opposite velocity $v_{io}$ ($v_{io}<0$).
Therefore, the traffic jam moves with the velocity $v_{j}$ less than the velocity $v_{g}$ of jammed Keplerian gas particles.

We obtain the three velocities $v_{j}$, $v_{g}$ and $v_{io}$ to verify their relationship described above.
First, the velocity $v_{j}$ of the arm is calculated as follows.
Figs.\,\ref{fig:pv_1350_1360.eps}a and \ref{fig:pv_1350_1360.eps}b show
the snapshots of gas clump masses $m_{ij}^{t}$ at $t=1350$ and 1360 respectively.
In Fig.\,\ref{fig:pv_1350_1360.eps}, the arm is in counterclockwise rotation.
The rotation angle $\Delta\theta$ is given by about $\pi/60$ for a short period of time $\Delta t=10$ steps
and the radial distance $R$ by around 20.
Thus, the velocity $v_{j}$ is given by
\begin{equation}
v_{j}=\frac{\Delta\theta}{\Delta t}R
\sim\frac{\pi/60}{10}\times 20\sim 0.1.
\end{equation}
Second, the velocity $v_{g}$ of jammed Keplerian gas particles becomes $v_{g}\sim 0.2$ as already shown in Fig.\,\ref{fig:cm.eps}c.
Third, from Eq.\,\ref{eqn:vp}, the velocity $v_{io}$ is given by
\begin{equation}
v_{io}=v_{j}-v_{g}\sim -0.1.
\end{equation}
Thus, it is verified that the velocity $v_{io}$ is opposite to the velocity $v_{g}$ of jammed Keplerian gas particles, and the velocity $v_{g}$ of the arm is slower than the velocity $v_{g}$.

In the previous and present subsections, the dynamic properties of the spiral arm have been studied.
They agree with those of density waves \cite{Lin-Shu} well, except that the spiral arm is transient, as discussed in the following subsection.

\subsection{Disappearance of a spiral arm}
\label{Disappearance of a spiral arm}

We now evaluate simply and directly the disappearance of a spiral arm based on the ``light-in and heavy-out'' approach, which considers the dynamic nature of patterns.
This evaluation shows that the spiral arm is transient as presented in Fig.\,\ref{fig:lmf.eps}.
In addition, we estimate the remaining lifetime of the spiral arm by using an approximate formula
which is immediately derived from the ``light-in and heavy-out'' approach.

The ``light-in and heavy-out'' is the gas flow rate difference between a light (i.e., low density) inflow into and heavy (i.e., high density) outflow from the traffic jam in spiral arms.
Here the gas inflow rate (and also outflow rate) is given by the product of its density and speed.

In Fig.\,\ref{fig:cm.eps}b,
the outflow gas density $m_{out}=\tilde{m}_{34}$ (the red circle in Fig.\,\ref{fig:cm.eps}b) at the front of the traffic jam is about two times higher than the inflow gas density $m_{in}=\tilde{m}_{20}$ (the blue circle in Fig.\,\ref{fig:cm.eps}b) at the tail.
This is because gas particles are jammed and compressed, so that they become massive gas clumps in the jam,
which we interpret as the formation of stars \cite{GalaDy} (or possibly even black holes \cite{Rana}).

In Fig.\,\ref{fig:cm.eps}c,
the outflow gas speed $v_{out}=|\tilde{\myvec{v}}_{34}|$ (the red circle in Fig.\,\ref{fig:cm.eps}c) at the jam front is slower than the inflow gas speed $v_{in}=|\tilde{\myvec{v}}_{20}|$ (the blue circle in Fig.\,\ref{fig:cm.eps}c) at the jam tail,
but the ratio $v_{out}/v_{in}$ is almost one. This is because gas particles are in Keplerian motion.

As a consequence of the above discussion, the gas outflow rate $m_{out}v_{out}$ at the jam front is more than the gas inflow rate $m_{in}v_{in}$ at the jam tail.
This gas flow rate difference $m_{out}v_{out}-m_{in}v_{in}$ decreases the mass $m_{jam}$ of the jam, and leads to the disappearance of the traffic jam.
Here we call this astronomical gas flow rate difference the ``light-in and heavy-out''.

Assuming that the gas flow rate difference $m_{out}v_{out}-m_{in}v_{in}$ is constant,
the remaining lifetime $t_{d}$ of the spiral arm is immediately given by the following approximate formula, without calculating its pattern speeds:
\begin{equation}
\label{eqn:td}
t_{d}=\frac{m_{jam}-m_{ofs}}{m_{out}v_{out}-m_{in}v_{in}}.
\end{equation}
The mass $m_{jam}$ ($m_{jam}=\sum_{i=21}^{34}\tilde{m}_{i}\times 1$ in Fig.\,\ref{fig:cm.eps}b) decreases gradually by the gas flow rate difference $m_{out}v_{out}-m_{in}v_{in}$.
It reaches the mass $m_{ofs}$ with the density $m_{out}$ ($m_{ofs}=\sum_{i=21}^{34}m_{out}\times 1$ with the red rectangle in Fig.\,\ref{fig:cm.eps}b), and finally the spiral arm disappears.

In order to verify the approximate formula Eq.\,\ref{eqn:td},
we estimate the remaining lifetime of the spiral arm at $t=1360$ shown in Fig.\,\ref{fig:cm.eps}.
The remaining lifetime $t_{d}$ is given by
\begin{equation}
\label{eqn:td1360}
t_{d}=\frac{1.5\times 10^{-5}-0.8\times 10^{-5}}{1.2\times 10^{-7}-0.8\times 10^{-7}}=175.
\end{equation}
Eq.\,\ref{eqn:td1360} indicates that the spiral arm disappears at about $t=1360+t_{d}=1535$, which is indeed confirmed with the spiral arm almost disappearing in the snapshot at $t=1520$ in Fig.\,\ref{fig:lmf.eps}.

In Eq.\,\ref{eqn:td}, we may use the average gas speed at radial distance $R$ (for example, the Keplerian speed $\propto \sqrt{1/R}$ or the rotation speed \cite{Kuno2}) instead of the gas speeds $v_{in}$ and $v_{out}$.
This modification would be useful when we apply the approximate formula to observations, since gas velocity (or speed) is difficult to measure \cite{Kuno2}.

\section{Summary and discussion}
\label{summary}

In this paper, we have studied in detail the dynamic properties of spiral arms simulated by the CML for astronomical objects,
and especially have discussed the disappearance of spiral arms.
The dynamic properties were shown from the following three aspects in the dynamics of traffic jams.
First, spiral arms are astronomical traffic jams formed by jammed Keplerian gas around the central star.
Second, the motion of the spiral arm is given as the movement of the traffic jam,
which results from both a gas inflow into and outflow from the jam.
Third, the disappearance of the spiral arm is simply and directly evaluated by the gas flow rate difference between the light inflow and heavy outflow in the jam,
which we have called the ``light-in and heavy-out'' approach.

In addition, 
we have proposed an approximate formula for the remaining lifetime of spiral arms
based on the ``light-in and heavy-out'' approach.
The approximate values obtained are consistent with the remaining lifetimes in the CML simulations.
This approximate formula can be generally applied to the observational data of spiral arms
(for example, gas density and velocity profiles in the streaming motions \cite{GalaDy,Kuno2}) independently of the detailed dynamics
of spiral galaxies or protoplanetary disks
by assuming that spiral arms are of the nature of traffic jams.
We will soon report elsewhere such application to the observational data of spiral galaxy M51 \cite{Kuno2}.

Finally, we briefly discuss the present study from the viewpoint of high-dimensional dynamical systems.
As described in Section \ref{Dynamic properties of a spiral arm}, the system of gas clumps transitions to an ordered state with the formation and disappearance of spiral arms, via a disordered state with the gas ejection from the central star. Moreover, the transition between ordered and disordered states is repeated over and over again for a long time, as reported in \cite{Nozawa}. Several observations and simulations have been reported which suggest the gas ejection from the central star and its repetition, for example, in studies of outbursts from an active galactic nucleus in galaxies \cite{Jones,Schlegel} and bursts from a protostar in protoplanetary disks \cite{Vorobyov,Vorobyov2}. We consider that this transition and repetition behavior is related to chaotic itinerancy in high-dimensional dynamical systems \cite{Kaneko,Tsuda}, especially astronomical chaotic itinerancy \cite{Konishi} based on the nonlinear dynamics of the central star whose four massive gas clump elements interact gravitationally.
We will introduce a new approach to describe the change in dynamic behavior of the central star from the dynamical systems theory such as bifurcation, chaos and chaotic itinerancy, in future work to explore the origin of spiral arms.

\section*{Acknowledgement}\vspace{-0.2cm}

The author would like to thank T. Deguchi for critical reading and suggestive comments on the manuscript.
The author also would like to thank M. Morikawa for extensive discussions on the density waves.
The author extends her sincere gratitude to K. Kaneko and I. Tsuda for helpful discussions and warm encouragement.
Finally, the author is grateful to an anonymous referee for carefully reading the manuscript and providing valuable comments which helped to improve the content.
This work was supported by JSPS KAKENHI Grant Number JP20J12074 and Ochanomizu University Nagase Research Scholarship.

\let\doi\relax

\vspace{-0.7cm}

\appendix

\section{Area-weighted average of field variables}\vspace{-0.2cm}
\label{Area-weighted average of field variables}

We define the field variables at any position $\myvec{r}$
as the area-weighted average of the field variables on the lattice.
The area-weighted average field variable $\tilde{a}^{t}(\myvec{r})$ is given by the following equation
with the field variables $a_{ij}^{t}$ at the four nearest neighboring lattice points of the position $\myvec{r}$:
\begin{eqnarray}
\label{eq:fvxy}
\lefteqn{\tilde{a}^{t}(\myvec{r})=\tilde{a}^{t}(x,y)}
\hphantom{=\left\{1-\left(x-\lfloor x\rfloor\right)\right\}\left\{1-\left(y-\lfloor y\rfloor\right)\right\}a_{\lfloor x\rfloor\lfloor y\rfloor}^{t}+\left(x-\lfloor x\rfloor\right)\left\{1-\left(y-\lfloor y\rfloor\right)\right\}a_{\lfloor x\rfloor +1\lfloor y\rfloor}^{t}}
\nonumber\\
=\left\{1-\left(x-\lfloor x\rfloor\right)\right\}\left\{1-\left(y-\lfloor y\rfloor\right)\right\}
a_{\lfloor x\rfloor\lfloor y\rfloor}^{t}
+\left(x-\lfloor x\rfloor\right)\left\{1-\left(y-\lfloor y\rfloor\right)\right\}a_{\lfloor x\rfloor +1\lfloor y\rfloor}^{t}\nonumber\\
\lefteqn{+\left\{1-\left(x-\lfloor x\rfloor\right)\right\}\left(y-\lfloor y\rfloor\right)a_{\lfloor x\rfloor\lfloor y\rfloor +1}^{t}
+\left(x-\lfloor x\rfloor\right)\left(y-\lfloor y\rfloor\right)a_{\lfloor x\rfloor +1\lfloor y\rfloor +1}^{t},}
\hphantom{=\left\{1-\left(x-\lfloor x\rfloor\right)\right\}\left\{1-\left(y-\lfloor y\rfloor\right)\right\}a_{\lfloor x\rfloor\lfloor y\rfloor}^{t}+\left(x-\lfloor x\rfloor\right)\left\{1-\left(y-\lfloor y\rfloor\right)\right\}a_{\lfloor x\rfloor +1\lfloor y\rfloor}^{t}}
\end{eqnarray}
where the coefficient of each field variable $a_{ij}^{t}$ gives the area.

\end{document}